# Measuring the Impact of Spectre and Meltdown

Andrew Prout, William Arcand, David Bestor, Bill Bergeron, Chansup Byun, Vijay Gadepally, Michael Houle,
Matthew Hubbell, Michael Jones, Anna Klein, Peter Michaleas, Lauren Milechin, Julie Mullen, Antonio Rosa,
Siddharth Samsi, Charles Yee, Albert Reuther, Jeremy Kepner

MIT Lincoln Laboratory, Lexington, MA, U.S.A.

*Abstract*—The Spectre and Meltdown flaws in modern microprocessors represent a new class of attacks that have been difficult to mitigate. The mitigations that have been proposed have known performance impacts. The reported magnitude of these impacts varies depending on the industry sector and expected workload characteristics. In this paper, we measure the performance impact on several workloads relevant to HPC systems. We show that the impact can be significant on both synthetic and realistic workloads. We also show that the performance penalties are difficult to avoid even in dedicated systems where security is a lesser concern.

## I. INTRODUCTION

Cybersecurity is a necessary component of all modern interconnected systems. High performance computing (HPC) systems present special challenges for cybersecurity as they are particularly sensitive to any performance impact of measures taken to enhance security.

Three new security vulnerabilities that were made public in January 2018 attack the design of most modern microprocessors. The variants of these vulnerabilities are known as Spectre (variant 1 and 2) [Kocher] and Meltdown (variant 3) [Lipp]. These vulnerabilities represent an entirely new class of attacks and have proven particularly difficult to mitigate. The mitigations that have been developed are advertised to have negative performance impacts. In this paper, we attempt to measure those impacts.

The organization of the rest of this paper is as follows. Section II describes the Spectre and Meltdown attacks in more detail. Section III describes the experimental environment. Section IV describes the benchmarks chosen to highlight the performance impact of the mitigations. Section V shows the performance results. Section VI summarizes the results.

## II. SPECTRE AND MELTDOWN VULNERABILITIES AND MIGATIONS

The Spectre and Meltdown vulnerabilities have been categorized into three variants.

### A. Variant 1: Spectre, Bounds Check Bypass

Spectre variant 1 allows for an attacker to cause the speculative execution engine to load data on the basis of an array index that is past the actual end of the array. While this action will never be committed by the speculative execution engine into the main flow of execution, this action can cause arbitrary memory outside of the array to be loaded into various caches. The presence of this data in cache can then be detected by other code, leaking information. A well-crafted attack can leak important information.

The primary area of concern for this variant has been javascript in web browsers and some other on-the-fly compiled code, such as the Berkley Packet Filter (BPF) Just-In-Time (JIT) compiler included in the Linux kernel. However, the BPF JIT is not enabled by default, lessening the concern for the Linux kernel. A small fix to validate array indexes provided to the Linux kernel originating from user-space was implemented; however, it is not tested for this paper because of its expected impact being too small for practical measurement and the lack of a flag to disable it.

### B. Variant 2: Spectre, Branch Target Injection

Spectre variant 2 allows for an attacker to cause the speculative execution engine to mispredict an indirect jump to a branch of code. It does this by performing various operations that cause the branch target predictor to speculate a branch to an attacker-chosen location. This misprediction can then be used to execute a Spectre variant 1 attack in places not normally accessible to such an attack.

Fixing this flaw requires a combination of several new processor instructions collectively known as Indirect Branch Control (IBC), delivered to existing chips via microcode updates, and further software changes to prevent the speculative execution engine from taking a mispredicted branch. The software fixes utilize these new IBC processor instructions in appropriate places and change the instructions used to implement indirect branches. This latter change, which has become known as retpolines, changes an indirect branch from being implemented with the `JMP` or `CALL` instruction to setting the `RSB` register and issuing a `RET` instruction, traditionally used to return to the previous location after a `JMP` or `CALL`. This nontraditional method of performing an indirect branch bypasses the indirect branch predictor entirely, as `RET` instructions were never envisioned to be used in this manner. Both of these changes require recompilation of affected software, potentially a significant barrier for closed-source or legacy code.

### C. Variant 3: Meltdown, Rogue Data Cache Load

Meltdown (known as variant 3) allows for an attacker to cause the speculative execution engine to read a memory

This material is based upon work supported by the Assistant Secretary of Defense for Research and Engineering under Air Force Contract No. FA8721-05-C-0002 and/or FA8702-15-D-0001. Any opinions, findings, conclusions or recommendations expressed in this material are those of the author(s) and do not necessarily reflect the views of the Assistant Secretary of Defense for Research and Engineering.

address not normally accessible, in advance of the permission check for the read operation. The speculative execution engine will discard the read when it later discovers the result of the permission check and prior to marking the speculative execution as complete. However, side effects of this speculative read, especially the loading of the speculatively read memory address into cache, can be detected through various side channels.

The near-term fix to this flaw has been to separate the page tables of user-space and kernel-space. Previously kernel memory was mapped into all user-space processes for performance, relying on the memory permission flags to prevent inappropriate access. Splitting the kernel memory space into its own page table prevents speculative reads from user-space being able to target kernel memory, at the cost of an additional page table switch during every transition from user-space to kernel-space and back (every syscall) and additional overhead when coping user memory to kernel functions. This change has become known as Kernel Page Table Isolation (KPTI).

GRSecurity and the associated PAX project has long had a feature that "hardened" and performed additional checks on memory access between user-space and kernel-space. This feature is called UDEREF, for user-space dereference. The mainline kernel's KPTI feature was merged with these existing protections in GRSecurity-enabled kernels and cannot be separately enabled or disabled.

## III. EXPERIMENTAL ENVIRONMENT

The MIT Lincoln Laboratory Supercomputing Center provides a high performance computing platform to over 1000 users at MIT, and is heavily focused on highly iterative interactive supercomputing and rapid prototyping workloads [Reuther 2004, Bliss]. As part of our mission to deliver new and innovative technologies and methods enabling scientists and engineers to quickly ramp up the pace of their research by leveraging big compute and big data storage assets, we have built the MIT SuperCloud [Reuther 2013], a fusion of the four large computing ecosystems: supercomputing, enterprise computing, big data and traditional databases into a coherent, unified platform. The MIT SuperCloud has spurred the development of a number of cross-ecosystem innovations in high performance databases [Byun, Kepner 2014], database management [Prout 2015], data protection [Kepner 2014b], database federation [Kepner 2013, Gadepally], data analytics [Kepner 2012] and system monitoring [Hubbell].

### A. Hardware Platform

All the experiments described in this paper were performed on the TX-Green Supercomputer at MIT Lincoln Laboratory.

The TX-Green Supercomputer used to perform these environmental simulations is a petascale system that consists of a heterogeneous mix of AMD, Intel, and Knights Landing-based servers connected to a single, non-blocking 10 Gigabit Ethernet Arista DCS-7508 core switch. All of the compute nodes used in these performance tests were Intel Xeon E5-2683 v3 (Haswell) servers with 256 GB of system RAM. The Lustre [Braam] central storage system uses a 10 petabyte Seagate ClusterStor CS9000 storage array that is directly connected to the core switch, as is each individual cluster node. This architecture provides high bandwidth to all the nodes and the central storage, and is depicted in Figure 1.

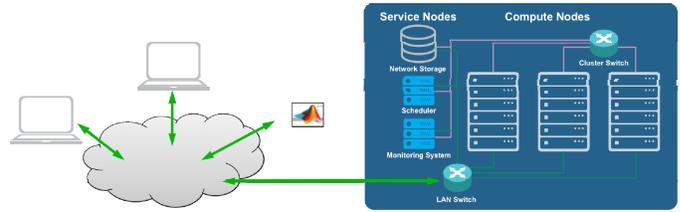

Figure 1. Architecture of the MIT SuperCloud system. Users connect to the system over either a local area network or a wide area network. At the time of connection, their system joins the MIT SuperCloud and can act as a compute node in order to run parallel programs interactively. The centerpiece of the MIT SuperCloud is several file systems (Seagate, DDN, Dell, Hadoop, and Amazon S3) running on several different network fabrics (10 GigE, InfiniBand, OmniPath).

### B. Software Environment

Our tests were performed on the MIT SuperCloud platform running GridOS 26, a derivative of Red Hat Fedora 26. The systems run Linux kernel version 4.4.131 and Lustre client 2.10.3. The tests were performed both with and without GRSecurity. In all, six different compilations of the Linux kernel and 10 configurations were tested.

The baseline configuration had all of the recent Linux kernel changes to address Meltdown or Spectre set to disabled and microcode revision 0x2e, which does not include the IBC updates. The first test configuration had PAGE_TABLE_ISOLATION configuration option enabled for the mainline kernel, and PAX_MEMORY_UDEREF configuration option enabled for the GRSecurity-enhanced kernel. The second test configuration was based on the first with RETPOLINE configuration option added. The third configuration was the same kernel as the second, but also had the system BIOS updated to the latest release that included microcode revision 0x3c [Microcode]. The final configuration reverted to the baseline kernel, but with the BIOS updated with the 0x3c microcode. These configurations are show in Table 1.

These kernels were compiled with gcc 4.8.5-28.el7 for the baseline, test 1 and test 4. The kernel used for Test 2 and Test 3 was compiled with gcc 4.8.5-28.el7_5.1 to include the new compiler changes for retpolines.

|  | PTI/UDEREF | Retpoline | BIOS/Microcode |
|---|---|---|---|
| **Baseline** |  |  |  |
| **Test 1** | Yes |  |  |
| **Test 2** | Yes | Yes |  |
| **Test 3** | Yes | Yes | Yes |
| **Test 4** |  |  | Yes |

Table 1. Table of configurations tested for both mainline and GRSecurity-enhanced kernels.

## IV. BENCHMARK DESIGN

Several benchmarks were chosen to both highlight the maximum impact of the changes to address these vulnerabilities and evaluate their impact on realistic scientific workloads. In all cases, we performed the test multiple times, and took the fastest result. System disk caches were not cleared inbetween repeated runs, ensuring that the results are for a "hot cache" situation.

*A. Network Connection Establishment*

Establishing a TCP network connection is a fundamental component of almost all uses of a network computer. In scientific computing, these connections are often used from within an MPI library or other framework to communicate among the systems participating in a distributed computation. For this test, we decided to look at one of the worst-case usages of TCP: transmitting a single byte. Most real-world uses of TCP will not behave this way and will keep an established TCP connection open to transfer more data. However, the heavy reliance on the operating system kernel during TCP connection establishment allows us to focus on the impact of the KPTI modification to address the Meltdown vulnerability, which is known to slow down transitions between user-mode and kernel-mode operation required by syscalls.

We performed this test both with and without the User-Based Firewall [Prout 2016] for controlling access between nodes. We used netcat on both the client and server systems which were both configured with the appropriate kernel for each test. While the impact of the User-Based Firewall itself had already been previously profiled, it adds a significant number of syscalls to the process of establishing the connection, which may be adversely effected by the KPTI modification.

*B. Disk Access*

Reading and writing to disk are fundamental operations of computer programs. For this test, we decided to look at one of the worst-case usages of disk access: writing one byte at a time. There are many ways to optimize disk utilization, both generally and specific to scientific computing or high performance file systems. However, optimization requires the application author to have done so, or be willing to do so. A significant number of applications have not done so. It is often impractical to modify these applications for a variety of reasons: they are closed-source, part of a large open-source project that would require significant investment to modify, or experimental code that is not expected to be used often and not worth the developer's time investment. This test again allows us to focus on the impact of the KPTI modification which is known to slow down transitions between user-mode and kernel-mode operation required by syscalls.

We performed this test on both the local filesystem backed by traditional spinning disk and the high performance parallel network filesystem running Lustre. We used the "dd" program to write 10 MB of data from the "zero device" to a new file for each test. While previous testing has highlighted the performance that is possible for optimized access to Lustre [Jones], the syscall heavy nature of writing one byte at a time was expected to be adversely affected by the KPTI modification.

*C. Computationally intensive code*

We tested two computationally intensive workflows, pMatlab and Keras with Tensorflow.

MATLAB® is commonly used for scientific computing and pMatlab (parallel Matlab) provides the capability to perform that computation in parallel [Bliss 2007]. pMatlab is used extensively at MIT Lincoln Laboratory both with MATLAB® and the open-source GNU Octave project.

pMatlab comes with several examples as part of the distribution. For this test, we chose to run the pBlurimage example with 16 worker processes. We ran these in local mode on a single node to eliminate the variability of network access and the HPC scheduler. These tests were run with matlab version 2017B.

The rise of effective machine learning has resulted in the development of many deep learning frameworks and libraries such as Caffe, TensorFlow, Torch, scikit-learn, Theano, DIGITS, and many more. While many of these often lean heavily on special-purpose hardware (e.g., GPUs or TPUs) and software (e.g., CUDA and cuBLAS) that relieve the primary processor of most of the computational load of these tasks and eliminate many instances of user-to-kernelspace context switching, the architecture of newer vector processors, such as the Intel Xeon Phi (Knight's Landing), enables competitive machine learning rates without the use of an additional coprocessor.

Our machine learning test trains a simple convolutional neural network (CNN) on the MNIST database, a large dataset consisting of handwritten digits frequently used in machine learning training, applying the Keras deep-learning framework atop a TensorFlow 1.6 backend. The TensorFlow, Keras, and Numpy packages used here were sourced from the Intel Distribution for Python as provided in the Anaconda Cloud. These packages leverage Intel's Math Kernel Library for Deep Neural Networks (MKL-DNN), a library providing highly parallel and vectorized mathematical functions that improve the performance of machine learning applications on x86 processors.

V. PERFORMANCE RESULTS

The results of our performance testing are shown in Figures 2, 3 and 4.

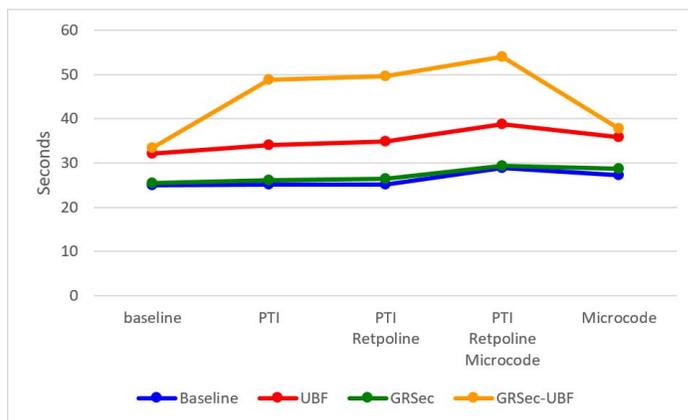

Figure 2. Time to establish 1,000 TCP connections and send one byte each via netcat.

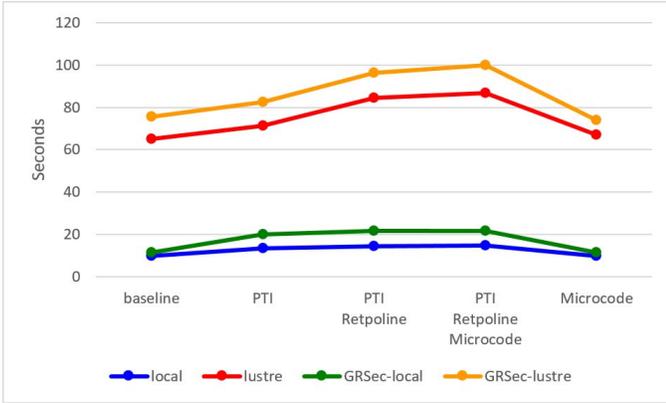

Figure 3. Time to copy 10 MB of data from /dev/zero to a new file with a block size of 1 byte using dd.

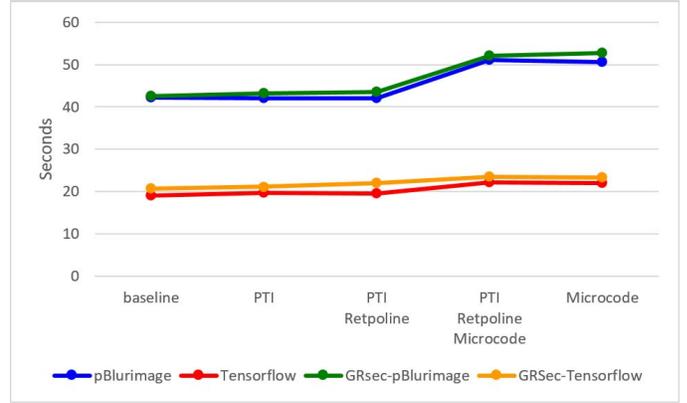

Figure 4. Time to execute the pBlurimage example with 16 local computation threads and Keras Tensorflow on the MNIST database with a batch size of 512.

Figure 2 shows the results of our tests on network connection establishment. The impact is much higher with the User-Based Firewall than without. This result is not unexpected as the firewall adds a significant number of user-kernel mode transitions to the process of accepting an inbound TCP connection. However, the impact of the UDEREF feature in GRSecurity is significantly greater than the impact of KPTI mitigation in the mainline kernel. With all mitigations enabled, the mainline kernel is slowed down by approximately 15% without and 21% with the User-Based Firewall. The GRSecurity-enabled kernel is also slowed down by 15% without, but 67% with the User-Based Firewall.

Figure 3 shows the results of our tests on disk access. The impact on local disk access is much higher than on Lustre access. This is likely caused by Lustre being slower overall for the small block size chosen due to the required network round-trips, hiding some of the impact of the mitigations. With all mitigations enabled, the mainline kernel is slowed down by approximately 50% on local disk and 33% on Lustre. The GRSecurity-enabled kernel is slowed down by 90% on local disk and 33% on Lustre.

Figure 4 shows the results of our tests on computationally intensive code. As expected, the kernel changes appear to have little effect on these benchmarks as they make few requests for kernel services (syscalls). However, a noticeable slowdown was seen with the microcode updated, which inspired Test 4 to be added. As shown in the Test 4 configuration, the performance impact remained even when all advertised mitigations were deactivated simply by having a CPU running the new microcode. These slowdowns were measured at 21% for pMatlab and 16% for TensorFlow for the baseline kernel with all mitigations (Test 3) and 19% and 15% respectively with just the microcode (Test 4).

## VI. SUMMARY

The Spectre and Meltdown vulnerabilities represent a novel class of attacks. The original announcement of these vulnerabilities aptly stated that Spectre will haunt us for quite some time. New attacks within these classes are expected to be discovered requiring additional nontrivial mitigations.

Our testing shows that the mitigations for these vulnerabilities can have significant impacts. Worse yet, for the microcode update the impacts are present even with all software mitigations turned off. This means that in systems where performance is more important than security, such as in a closed non-internet-connected system with a dedicated userbase, it is not easy to maintain that performance if a BIOS update is needed to address some other issue with the system.

Our testing also shows a significant deviation between the mainline and GRSecurity-enabled kernels. The majority of the deviation for the GRSecurity-enabled results appears with the UDEREF mitigation. While UDEREF is advertised to be comparable to KPTI, our testing suggests it is not.

The choice of unoptimized benchmarks for network connection establishment and disk access was intended to show the worst-case scenario. While well-optimized code will experience significantly less impact than the worst case, we find that we run a vast amount of unoptimized code. MIT Lincoln Laboratory Supercomputing Center focuses on interactive supercomputing and rapid prototyping. These workloads are often unoptimized because of their one-time use nature or the immaturity of the project they support. It has not previously been worthwhile to invest the staff time toward correcting this, given the size of the performance penalties involved. However, the wisdom of this decision is predicated on knowing the size of the penalty involved.

REFERENCES

[Bliss] N. T. Bliss, R. Bond, J. Kepner, H. Kim, and A. Reuther, "Interactive grid computing at Lincoln Laboratory," Lincoln Laboratory Journal, vol. 16, no. 1, p. 165, 2006.

[Braam] P. J. Braam, The Lustre Storage Architecture. Cluster File Systems, Inc., October 2003.

[Byun] C. Byun, W. Arcand, D. Bestor, B. Bergeron, M. Hubbell, J. Kepner, A. McCabe, P. Michaleas, J. Mullen, D. O'Gwynn, A. Prout, A. Reuther, A. Rosa, and C. Yee, "Driving Big Data with Big Compute," in High Performance Extreme Computing Conference (HPEC), IEEE, 2012.

[Gadepally] V. Gadepally, J. Kepner, W. Arcand, D. Bestor, B. Bergeron, C. Byun, L. Edwards, M. Hubbell, P. Michaleas, J. Mullen, A. Prout, A. Rosa, C. Yee, and A. Reuther, "D4M: Bringing associative arrays to database engines," in High Performance Extreme Computing Conference (HPEC), IEEE, 2015.

[Hubbell] M. Hubbell, A. Moran, W. Arcand, D. Bestor, B. Bergeron, C. Byun, V. Gadepally, P. Michaleas, J. Mullen, A. Prout, A. Reuther, A. Rosa, C. Yee, and J. Kepner, "Big Data strategies for Data Center

Infrastructure management using a 3D gaming platform," in High Performance Extreme Computing Conference (HPEC), IEEE, 2015.

[Jones] M. Jones, J. Kepner, W. Arcand, D. Bestor, B. Bergeron, V. Gadepally, M. Houle, M. Hubbell, P. Michaleas, A. Prout, A. Reuther, S. Samsi, P. Monticiollo, "Performance Measurements of Supercomputing and Cloud Storage Solutions," IEEE High Performance Extreme Computing (HPEC) Conference, Sep 12-14, 2017, Waltham, MA.

[Kepner 2012] J. Kepner, W. Arcand, W. Bergeron, N. Bliss, R. Bond, C. Byun, G. Condon, K. Gregson, M. Hubbell, J. Kurz, A. McCabe, P. Michaleas, A. Prout, A. Reuther, A. Rosa, and C. Yee, "Dynamic Distributed Dimensional Data Model (D4M) database and computation system," in 2012 IEEE International Conference on Acoustics, Speech and Signal Processing (ICASSP), pp. 5349–5352, IEEE, 2012.

[Kepner 2013] J. Kepner, C. Anderson, W. Arcand, D. Bestor, B. Bergeron, C. Byun, M. Hubbell, P. Michaleas, J. Mullen, D. O'Gwynn, A. Prout, A. Reuther, A. Rosa, and C. Yee, "D4M 2.0 schema: A general purpose high performance schema for the Accumulo database," in High Performance Extreme Computing Conference (HPEC), IEEE, 2013.

[Kepner 2014a] J. Kepner, W. Arcand, D. Bestor, B. Bergeron, C. Byun, V. Gadepally, M. Hubbell, P. Michaleas, J. Mullen, A. Prout, A. Reuther, A. Rosa, and C. Yee, "Achieving 100,000,000 database inserts per second using accumulo and d4m," in High Performance Extreme Computing Conference (HPEC), IEEE, 2014.

[Kepner 2014b] J. Kepner, V. Gadepally, P. Michaleas, N. Schear, M. Varia, A. Yerukhimovich, and R. K. Cunningham, "Computing on masked data: a high performance method for improving big data veracity," in High Performance Extreme Computing Conference (HPEC), IEEE, 2014.

[Kocher] Kocher, Paul and Genkin, Daniel and Gruss, Daniel and Haas, Werner and Hamburg, Mike and Lipp, Moritz and Mangard, Stefan and Prescher, Thomas and Schwarz, Michael and Yarom, Yuval, "Spectre Attacks: Exploiting Speculative Execution", arXiv:1801.01203 [cs.CR], Jan. 2018.

[Lipp] M. Lipp, M. Schwarz, D. Gruss, T. Prescher, W. Haas, S. Mangard, P. Kocher, D. Genkin, Y. Yarom, and M. Hamburg, "Meltdown", arXiv:1801.01207 [cs.CR], Jan. 2018.

[Microcode] Microcode Revision Guidance, March 6 2018, <https://newsroom.intel.com/wp-content/uploads/sites/11/2018/03/microcode-update-guidance.pdf>.

[Prout 2015] A. Prout, J. Kepner, P. Michaleas, W. Arcand, D. Bestor, B. Bergeron, C. Byun, L. Edwards, V. Gadepally, M. Hubbell, J. Mullen, A. Rosa, C. Yee, and A. Reuther, "Enabling On-Demand Database Computing with MIT SuperCloud Database Management System," IEEE High Performance Extreme Computing (HPEC) Conference, Sep 15-17, 2015, Waltham, MA.

[Prout 2016] A. Prout, W. Arcand, D. Bestor, D. Bergeron, C. Byun, V. Gadepally, M. Hubbell, M. Houle, M. Jones, P. Michaleas, L. Milechin, J. Mullen, A. Rosa, S. Samsi, A. Reuther, and J. Kepner, "Enhancing HPC Security with a User-Based Firewall," IEEE High Performance Extreme Computing (HPEC) Conference, Sep 13-15, 2016, Waltham, MA.

[Reuther 2004] A. Reuther, T. Currie, J. Kepner, H. G. Kim, A. McCabe, M. P. Moore, and N. Travinin, "LLGrid: Enabling on-demand grid computing with gridMatlab and pMatlab," tech. rep., MIT Lincoln Laboratory, 2004.

[Reuther 2013] A. Reuther, J. Kepner, W. Arcand, D. Bestor, B. Bergeron, C. Byun, M. Hubbell, P. Michaleas, J. Mullen, A. Prout, and A. Rosa, "LLSuperCloud: Sharing HPC Systems for Diverse Rapid Prototyping," IEEE High Performance Extreme Computing (HPEC) Conference, Sep 10-12, 2013, Waltham, MA.